\documentclass[aps,prd,amssymb,final,nobibnotes,nofootinbib,superscriptaddress,showpacs]{revtex4}
\usepackage{graphicx}

\begin{document}
\title{The October 10, 1912 solar eclipse expeditions and the first attempt to measure light-bending by the Sun}
\par

\author{Lu\'{\i}s C. B. Crispino}
\email{crispino@ufpa.br} 
\affiliation{Faculdade de F\'{\i}sica,
Universidade Federal do Par\'a, 66075-110, Bel\'em, PA, Brasil}


\begin{abstract}
In 1911 Einstein proposed that light-bending by the Sun's gravitational field could be measured during 
a total solar eclipse. The first opportunity in which this measurement would be tried, was during the 
total solar eclipse of October 10, 1912.
We report about the expeditions sent to Brazil to observe this eclipse, including the one from the 
C\'ordoba Observatory, from Argentina, which aimed to measure this Einstein's effect.
\end{abstract}


\maketitle




During the V Amazonian Symposium in Physics~\cite{VASP:2019}, the centennial of the first experimental test 
of General Relativity was celebrated, in allusion to the well-known successful measurements performed during the total 
solar eclipse of May 29, 
1919~\cite{Crommelin:1919, DED:1920, Crelinsten:2006, Coles:2001, cl_ijmpd_2016, cl_pp_2016, 
c_ijmpd_2018, kenne_book2019, crisp_kenne2019}. 
What is less known is the fact that the first attempt to measure light deflection by the Sun's gravitational field 
was organized years before that, to be performed during the total 
solar eclipse of October 10, 1912~\cite{crisp_paola2020}. 
We report on the expeditions sent to Brazil, during this 1912 eclipse, including 
the one led by the American astronomer Charles Dillon Perrine, who at that time was the director of the 
Argentinean National Observatory, located in C\'ordoba, 
aiming to measure light deflection by the Sun's gravitational field.

%
%

\begin{figure}
\centering
\includegraphics[width=4.8in]{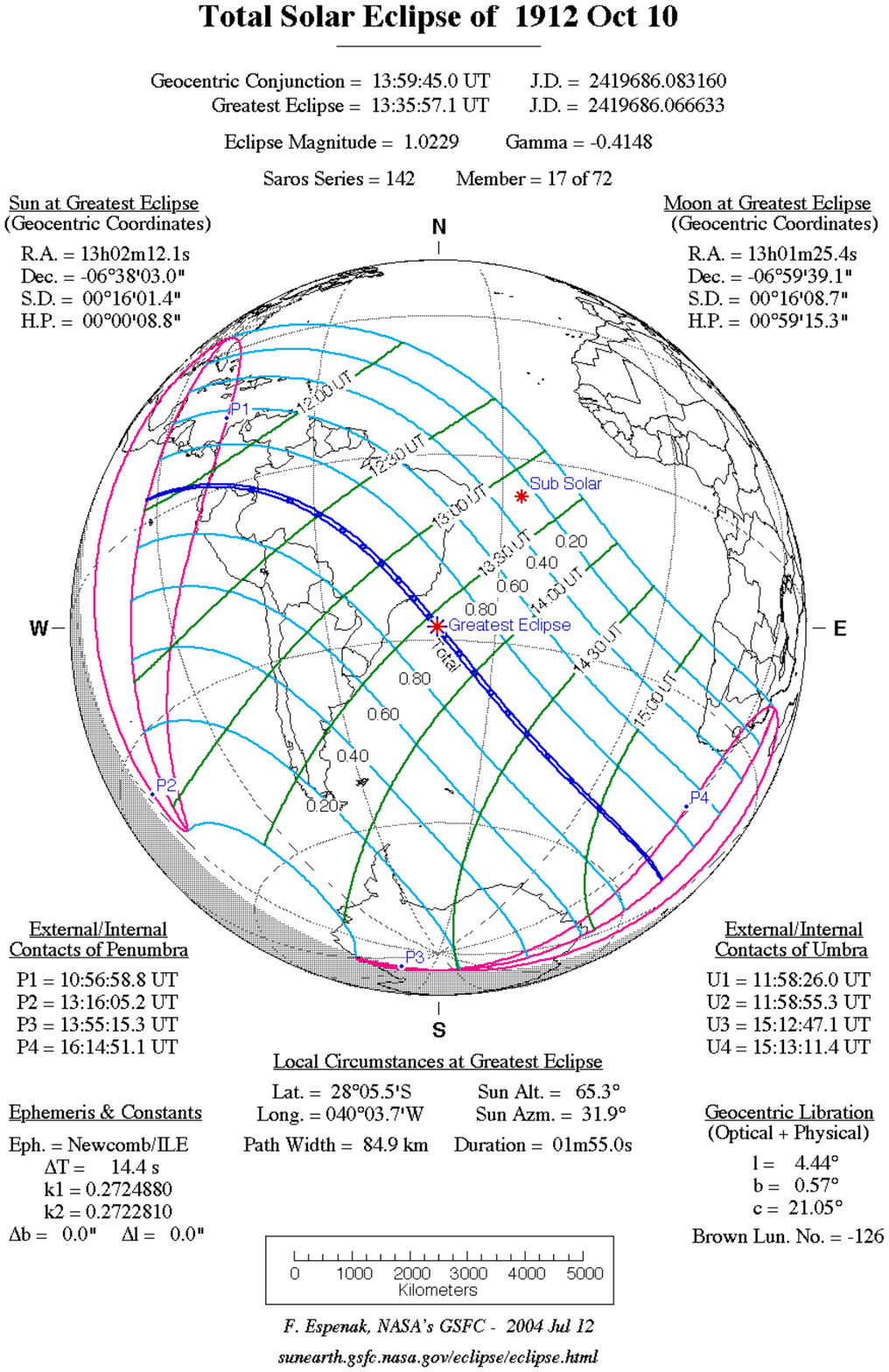}
\caption{Details of the October 10, 1912 total solar eclipse.
As it can be seen, Brazil had a privileged geographic location for the 
observation of this eclipse. 
Courtesy of NASA Eclipse Website~\cite{NASA} 
and Fred Espenak (Emeritus of NASA Goddard Space Flight Center), United States.}
\label{eclipse}
\end{figure}

Apart from regions in the Pacific and the Atlantic Oceans, 
the totality zone of the October 10, 1912 solar eclipse 
passed close to the border between Colombia, Equador and Peru, 
as well as across Brazil, including the states of Amazonas, 
Mato Grosso, Goi\'as, Minas Gerais, S\~ao Paulo 
and Rio de Janeiro (cf. Fig.~\ref{eclipse}).~\cite{NASA}

For the observation of this 1912 total solar eclipse, 
several expeditions came to Brazil.
Apart from the teams of the Brazilian National Observatory, 
there were commissions from 
British, French, American, Argentinean and Chilean institutions.
All of the foreign commissions arrived in Rio de Janeiro harbor, 
spent some time in town, and then traveled to the observation sites 
in the totality zone of the eclipse. 
The Brazilian Ministry of Finances waived all the custom duties  
related to the instruments transported by the 
scientists to observe the eclipse.~\cite{CdM:30aug1912}

\begin{figure}
\centering
\includegraphics[width=5.0in]{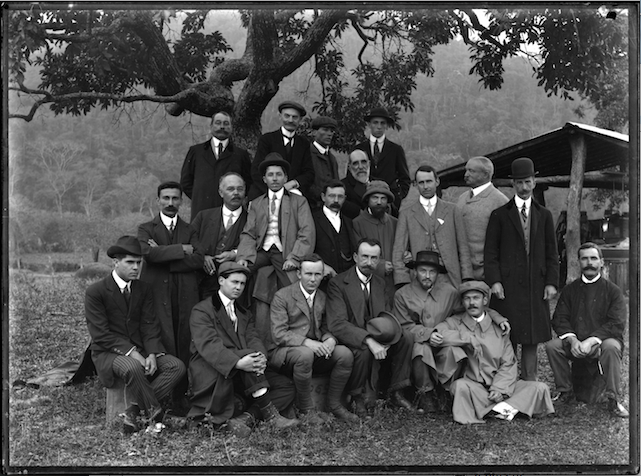}
\caption{Observers in Passa Quatro.
From left to right:
(i) First (front) row (seated):
Gualter Macedo Soares, Carlos Morize, James Henry Worthington, 
Henrique Charles Morize, Jaromir Kralicek, M\'ario Rodrigues de Souza 
e Augusto Soucasaux;
(ii) Second row: 
Domingos Fernandes da Costa, Theofilus Henry Lee, Alix Corr\^ea de Lemos, 
Charles Rundle Davidson, Milan Rastislav Stefanik, Arthur Stanley Eddington, 
John Jepson Atkinson and Ant\^onio Alves Ferreira da Silva;
(iii) Third row: 
Marc Ferrez;
(iv) Fourth row: 
Pierre Seux, Rodolpho Hess, Olyntho Couto de Aguirre e 
Leslie Andrews~\cite{AEpoca:03nov1912}. 
Courtesy of the {\it Biblioteca do Observat\'orio Nacional}, 
Rio de Janeiro, Brazil.}
\label{teams}
\end{figure}

\begin{figure}
\centering
\includegraphics[width=4.9in]{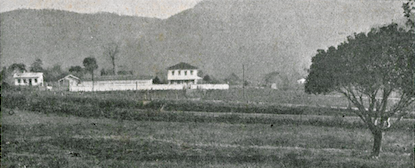}
\caption{View of Rodolpho Hess' farm, in Passa Quatro,
where the main 1912 Brazilian eclipse expedition, as well as the British and 
the French ones, set camp~\cite{FonFon:12out1912p36}.
Courtesy of the {\it Funda\c{c}\~ao Biblioteca Nacional}, 
Rio de Janeiro, Brazil.}
\label{hess_farm}
\end{figure}

\begin{figure}
\centering
\includegraphics[width=4.9in]{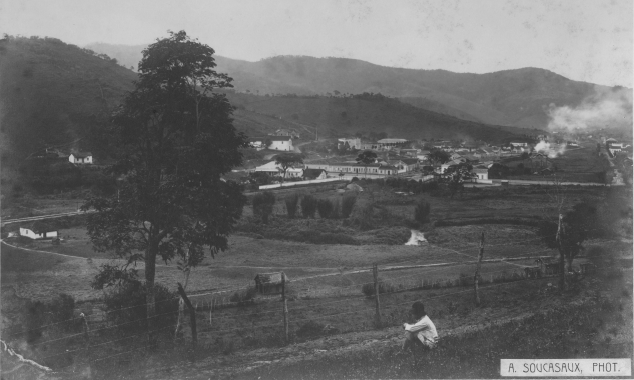}
\caption{General view of Passa Quatro. 
Photograph taken during the eclipse expedition of 1912.
Courtesy of the {\it Biblioteca do Observat\'orio Nacional}~\cite{ON}, 
Rio de Janeiro, Brazil.}
\label{PQ}
\end{figure}

The main Brazilian expedition, 
led by Henry (Henrique) Charles Morize (cf. Fig.~\ref{teams}), 
set camp in the {\it Bella Vista} farm (near the railway, surrounded by hills), 
owned by Rodolpho Hess (cf. Fig~\ref{teams} and~\ref{hess_farm}), 
distant about one kilometer (km) from the city of 
Passa Quatro (cf. Fig.~\ref{PQ}), 
in the Brazilian state of Minas Gerais.\footnote{
Not to be confused with the Brazilian cities of 
{\it Santa Rita do Passa Quatro}, in the state of S\~ao Paulo; nor 
with {\it S\~ao Miguel do Passa Quatro}, 
in the state of Goi\'as.}
In the same farm stayed the British and French expeditions.

The first observer from abroad to arrive in Brazil was the amateur astronomer 
James Henry Worthington (cf. Fig.~\ref{teams}), 
who reached Rio de Janeiro on September 2nd~\cite{ANoite:03set1912}. 
Pleased to be known as ``eclipse's hunter'', Worthington came to Brazil 
in his fourth eclipse expedition, the previous one being on April 17, 1912, 
in Portugal. In the day after his arrival in Rio, he visited the Brazilian National 
Observatory, located in {\it Morro do Castelo} (Castle Hill), 
meeting the director Morize.
Worthington was a rich man that decided to investigate, 
at his own expenses, scientific matters. 
He had special interest in eclipses. 
The equipment brought by Worthington to Brazil in 1912 
included two Steinheil photographic telescopes, also called coronographs, 
because of their use to take pictures of the solar corona. 
He also brought a Hilger quartz spectrograph, to analyze 
the spectrum of the solar corona during the eclipse. 
The coelostat (an instrument used to compensate Earth's rotation, 
keeping a fixed image in the cameras coupled to the 
instruments) used by Worthington had been constructed according to 
his own design (having three mirrors, one metallic and two composed 
of silvered glasses).~\cite{JdC:22set1912}

\begin{figure}
\centering
\includegraphics[width=4.9in]{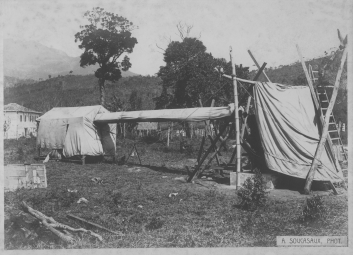}
\caption{French expedition installation in  Passa Quatro 
for the Mailhat telescope, with an objective lens of 10 m 
of focal distance. 
The coelostat is covered with canvas, in the right.
Inside the hut, in the left, is located the camera 
that captures the image of the solar corona. 
Courtesy of the {\it Biblioteca do Observat\'orio Nacional}, 
Rio de Janeiro, Brazil.}
\label{french}
\end{figure}

The expedition from the {\it Bureau des Longitudes}, 
located in France, 
was led by Milan Rastislav Stefanik, assisted by Jaromir Kralicek (cf. Fig.~\ref{teams}). 
Stefanik and Kralicek arrived in Rio de Janeiro on board 
of the French steamer {\it Amazone}, 
on September 10th.~\cite{ANoite:10set1912}
Their packing-cases were immediately liberated by the customs with 
the authorization of the Brazilian government. 
Stefanik gently brought with his load an equipment 
for the Brazilian National Observatory, that Morize had  
ordered for manufacturers in January 1912,
to be used for the eclipse observation by 
the Brazilian team~\cite{JdB:12set1912}. 
This equipment was a Mailhat telescope of 8 meters (m) of focal distance 
and 15 centimeters (cm) of aperture, coupled to a coelostat of the same 
manufacturer~\cite{Morize:13set1912}.
The French commission brought its own (bigger) 
Mailhat telescope of 10~m of focal distance, also coupled to a 
coelostat (cf. Fig.~\ref{french}), the whole set constituting 
a load of over two tons~\cite{FonFon:19out1912p29, JdC:10out1912}. 
Two other equipment brought by Stefanik were a 
spectrometer and a polarimeter of his own invention~\cite{JdB:09out1912}. 

Stefanik left Rio on September 11th, to visit the totality zone of the 
eclipse and to choose the best observation site. 
The first city he has chosen to visit was Cristina,~\cite{JdB:12set1912}
in the Minas Gerais state.
After a careful analysis, Stefanik decided to install his 
equipment in the city of Passa Quatro (cf. Fig.~\ref{french}), 
where Morize set the main Brazilian National Observatory camp.

\begin{figure}
\centering
\includegraphics[width=4.9in]{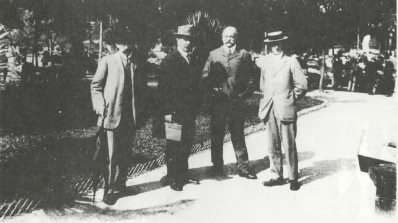}
\caption{Greenwich commission in the day 
of their arrival in Rio de Janeiro. 
From left to right: 
Charles R. Davidson, 
Henrique C. Morize, 
John J. Atkinson and 
Arthur S. Eddington.~\cite{Caffarelli:1980}
Courtesy of the {\it Ci\^encia e Cultura} Magazine, 
Brazil.}
\label{Pharoux}
\end{figure}

The Greenwich Observatory party arrived in Rio de Janeiro harbor on September 15th, 
on board of the ship {\it Arlanza}, and landed in the following day~\cite{mnras73-0386}. 
It was composed by Arthur Stanley Eddington  
(assistant astronomer of the Greenwich Observatory) and 
Charles Rundle Davidson
(chief computer of the Greenwich Observatory), 
led by the former (cf. Fig.~\ref{teams}).~\cite{mnras73-0386, Eddington:1913} 
John Jepson Atkinson~\footnote{
J. J. Atkinson accompained six British eclipse expeditions.
The first one was to Vads\"o in 1986, 
and the second one to Ovar, near Oporto, in 1900.
The third one was to Sumatra in 1901, and
the fourth one was to Tunis in 1905.
He participated in two eclipse expeditions in 1912. 
One was to St. Germains and the other to Brazil.
The one in Passa Quatro was his last eclipse 
expedition.~\cite{mnras_Obituary1925}}
joined the Greenwich expedition, accompanying
them from England. 
Morize and the chemist Theofilus Henry Lee (cf. Fig.~\ref{teams}), 
indicated by the Brazilian government to help the 
British party, met them in their arrival in the 
Pharoux quay~\cite{ANoite:16set1912} (cf. Fig.~\ref{Pharoux}). 

Eddington, Davidson, Lee and  Worthington 
left Rio in September 21st, 
taking the S\~ao Paulo express 
train to Cruzeiro (a city in the state of S\~ao Paulo). 
Their instruments had been sent in a previous train. 
In Cruzeiro the instruments had to be transferred  
to a narrow-gauge line, along which they were taken,  
through a 20 miles track, to Passa Quatro, 
where they arrived after a three-hours journey~\cite{Eddington:26set1912}. 

\begin{figure}
\centering
\includegraphics[width=4.9in]{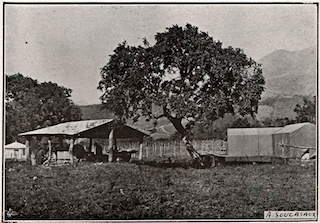}
\caption{Greenwich settlement (right) in  Passa Quatro. 
The covered area in the left served as a deposit for the materials 
of the British and the French commissions~\cite{FonFon:19out1912p29}.
Courtesy of the {\it Funda\c{c}\~ao Biblioteca Nacional}, 
Rio de Janeiro, Brazil.}
\label{English_Settlement}
\end{figure}

\begin{figure}
\centering
\includegraphics[width=4.9in]{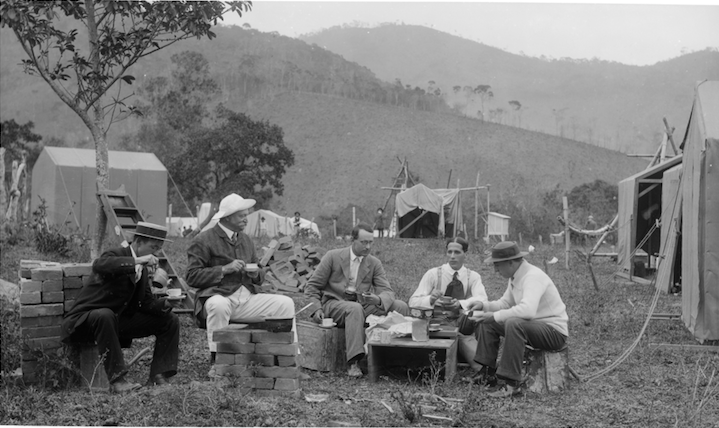}
\caption{Greenwich party having tea in  Passa Quatro.
From left to right:
Charles R. Davidson, John J. Atkinson, Arthur S. Eddington,
Olyntho C. de Aguirre and Leslie Andrews. 
The Greenwich huts, with their equipment, can be seen 
near the observers. 
Courtesy of the {\it Biblioteca do Observat\'orio Nacional}, 
Rio de Janeiro, Brazil.}
\label{tea}
\end{figure}

The British commission arrived in  
Passa Quatro on September 22nd. 
The following days were dedicated to transport 
the equipment to Rodolpho Hess' farm, 
erect the huts, mount the instruments (cf. Fig.~\ref{English_Settlement}), etc. 
Two volunteers from Rio de Janeiro, 
Olyntho Couto de Aguirre~\footnote{Olyntho C. de Aguirre 
was born in the Brazilian state of Esp\'irito Santo, 
and obtained his degree in Engineering in 
England.~\cite{Epoca:12out1912}}
and Leslie Andrews (cf. Fig.~\ref{teams}), joined the Greenwich team  
on October 3rd (cf. Fig.~\ref{tea}), 
to assist them in the work.~\cite{Eddington:1913}. 
There was a daily transportation (a locomotive engine) 
from the city of Passa Quatro, 
where the scientists were lodged, to the Hess' farm
(a distance of about one kilometer), 
where the instruments were positioned, leaving the city at 
8 am and coming back at 11 am, for lunch, 
returning 1 pm and back again at 6 pm~\cite{Eddington:26set1912}. 

The instruments brought by the Greenwich team were 
(i) a Thompson coronograph (to be operated by Eddington), 
to obtain large-scale photographs, 
with an object-glass of 9 inches (around 23~cm) aperture 
and 8 feet 6 inches focal length, 
in combination with a concave telephoto lens of 4 inches aperture 
and 16 inches focus 
(the total length of the instrument was 12 feet, with an equivalent 
focal length of 36 feet, or roughly 11~m), 
coupled to a 16 inch coelostat to reflect the light to the telescope; 
(ii) a six-inch refractor telescope (to be operated by Aguirre), 
of 7 feet focus, with green color filter (to investigate the presumed ``coronium''); 
(iii) a six-inch Cooke triplet (to be operated by Andrews), 
of 27 inches focus, 
also with green color filter; 
(iv) a spectrograph (to be operated by Davidson, assisted by Atkinson), 
made of quartz (used by Major Hills during the eclipse of 1898, and by Davidson during 
the eclipses of 1900, 1901, and 1905), 
coupled to a heliostat, to reflect the light to the spectrograph. 
The 6-inch telescope and the 6-inch triplet were mounted side by side, 
and fed by a second 16-inch coelostat.
The seconds during totality were supposed to be counted by Pierre Seux (cf. Fig.~\ref{teams}), 
who volunteered to that task, and did it during the rehearsals, which started 
on October 7th~\cite{mnras73-0386, Eddington:1913, JdC:22set1912}.

On board of the ship {\it Aragon}, 
the team from the Argentinean National Observatory, 
located at C\'ordoba, 
composed by its director Charles Dillon Perrine, 
the photographer Robert Winter, 
the engineer James Oliver Mulvey and 
the astronomer Enrique Chaudet, 
arrived in Rio de Janeiro on September 18th, 
bringing with them almost two tons of equipment.
In the following day Perrine visited the Brazilian National 
Observatory~\cite{ANoite:19set1912}.

\begin{figure}
\centering
\includegraphics[width=5.0in]{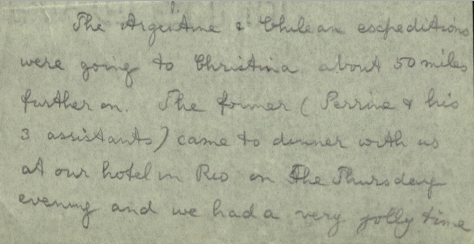}
\caption{Part of a letter from Arthur S. Eddington to his mother, 
Sarah Ann Eddington (dated from September 26, 1912), 
in which it is written: 
``The Argentine \& Chilean expeditions 
were going to Christina about 50 miles 
further on. The former (Perrine \& his
3 assistants) came to dinner with us 
at our hotel in Rio on the Thursday 
evening and we had a very jolly time.''~\cite{Eddington:26set1912} 
Courtesy of the Master and fellows of Trinity College, 
Cambridge, United Kingdom.}
\label{letter}
\end{figure}

The C\'ordoba party left Rio to the city of Cristina 
(located around 350 km away from Rio),  
on September 20th~\cite{ANoite:20set1912}. 
In the night before, the Greenwich and the 
C\'ordoba commissions dined together (cf. Fig.~\ref{letter}). 

\begin{figure}
\centering
\includegraphics[width=5.0in]{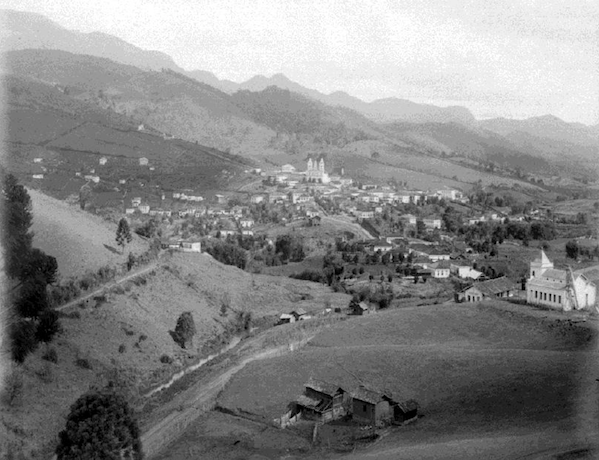}
\caption{General view of Cristina, in the Brazilian 
state of Minas Gerais.
Photo taken by the C\'ordoba Observatory team, 
during the 1912 eclipse expedition.
The  Perrine settlement is located in the right, 
attached to the big white building. 
Courtesy of the {\it Museo Astron\'omico del Observatorio Astron\'omico de C\'ordoba}, 
C\'ordoba, Argentina.}
\label{Cristina}
\end{figure}

\begin{figure}
\centering
\includegraphics[width=5.0in]{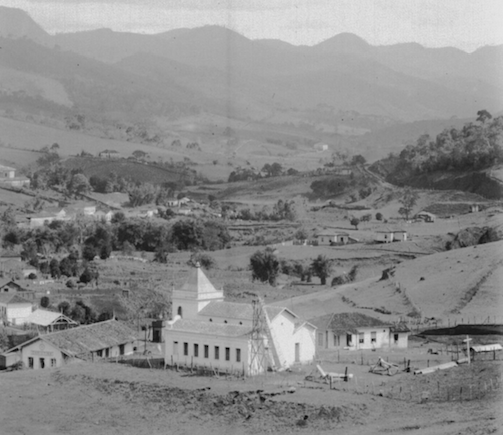}
\caption{View of the C\'ordoba Observatory 
settlement in Cristina.
Courtesy of the {\it Museo Astron\'omico del Observatorio Astron\'omico de C\'ordoba}, 
C\'ordoba, Argentina.}
\label{Cristina2}
\end{figure}

The  Brazilian National Observatory took charge of 
the transportation of the C\'ordoba team instruments 
from Rio to Cristina, arriving in the destination on 
September 24th.~\cite{Perrine_Report_1912}
The site chosen by Perrine in Cristina 
(cf. Figs.~\ref{Cristina} and~\ref{Cristina2})
to install his equipment, 
were the surroundings of a building 
(which, until that time, had never been used after its construction) 
located around one hundred meters away from 
the train station. 

\begin{figure}
\centering
\includegraphics[width=5.0in]{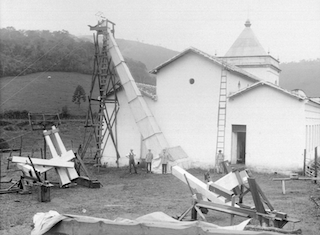}
\caption{Detail of the C\'ordoba settlement to 
observe the 1912 total solar eclipse.
The intramercurial cameras, used to test 
Einstein's prediction of light bending, 
can be seen in the left.  
In the bottom of the 12 m instrument, standing 
(from left to right) we see 
Robert Winter, 
James O. Mulvey and 
Enrique Chaudet. 
(The man in the bottom of the stairway could not 
be identified.)
Courtesy of the {\it Museo Astron\'omico del Observatorio Astron\'omico de C\'ordoba}, 
C\'ordoba, Argentina.}
\label{Cristina3}
\end{figure}

The equipment brought by the C\'ordoba Observatory 
team included about ten instruments (cf. Fig.~\ref{Cristina3}). 
The biggest one was a telescope of 12 m of focal distance. 
Among the instruments, there were two telescopes, 
coupled to cameras (the so-called intramercurial cameras), 
of 3 1/3~m of focal distance 
(loaned by William Wallace Campbell, 
director of the Lick Observatory, in the United States) 
for Perrine to try to measure 
the bending of the starlight passing close to the eclipsed Sun.
This effect, proposed by Albert Einstein in 1911~\cite{Einstein1911}, 
has been brought to the knowledge of Perrine 
by Erwin Finlay-Freundlich, 
assistant astronomer of the Berlin 
Observatory~\cite{crisp_paola2020, Lick, Perrine1923, PM:2007, PPMCOG:2018, paolantonio2019}. 

On September 25th, in the ship {\it Oronza}, more scientists 
from other American institutions arrived in Rio.~\cite{ANoite:25set1912}
From the Chilean Observatory, in Santiago, came 
Friedrich Wilhelm Ristenpart (director),~\footnote{
Ristenpart was working since 1908 as the 
director of the Santiago Observatory, 
had been hired, in Germany, for that purpose by the
Chilean government.
Already in the year of 1908, 
Ristenpart observed the Sun's eclipse in 
Corrientes~\cite{ANoite:25set1912}.
He had been elected Fellow of the Royal Astronomical 
Society on 14th June 1912.~\cite{Ristenpart}}
R\'omulo Grand\'on Moreno (assistant astronomer)
and
Richard W\"ust (mechanical engineer). 
From the Chilean Meteorological Institute, also in Santiago, came 
Walter Alfred Knoche (director) 
and 
Waldemar Trollund (mechanical engineer). 
From the La Plata Geophysics Department, in Argentina, came 
Jakob Johann Laub, 
together with his wife Ruth Elisa Wendt Laub. 
Morize himself went onboard of the {\it Oronza} to welcome 
these teams from the Chilean and Argentinean institutions.
%
%

The instruments of the Chilean Observatory included 
a photographic telescope of 16.20~cm (which had been used 
to observe the transit of Venus in front of the solar disk in 1874), 
a prismatic camera with a 10.45~cm lens, 
an equatorial Steinheil telescope of 10~cm aperture, 
specially bought from Germany 
for the 1912 eclipse, 
as well as a two small telescopes, 
sextants, chronometers and a 
chronograph.~\cite{AN_Ristenpart, Grandon}
Selenium cells were brought, to be coupled 
to optical instruments.

Knoche had been requested by Louis Agricola Bauer, 
director of the Department of Terrestrial Magnetism 
at the Carnegie Institution of Washington, to perform air-electrical 
measurements during the 1912 eclipse.~\cite{DTM, KL1}
Laub joined Knoche to comply with this Carnegie request. 
~\cite{KL1} 

Ristenpart, Knoche
and their teams set camp in the farm {\it Boa Vista} 
in the surroundings of Cristina. 
Knoche, Laub and Trollund were assisted in their observation site by Hermann Friedrich Albrecht von Ihering, 
the director of the {\it Museu Paulista} (located in the Brazilian city of S\~ao Paulo) and his wife Meta Buff;
by the senior teacher at the German School in Rio de Janeiro, Dr. Sh\"afer; 
as well as by Laub's wife, Ruth.~\cite{KL1}

\begin{figure}
\centering
\includegraphics[width=5.0in]{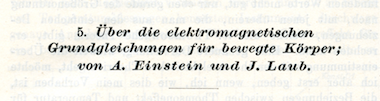}
\caption{Detail of the first page of the paper 
entitled {\it \"Uber die elektromagnetischen Grundgleichungen f\"ur bewegte K\"orper} 
(On the Fundamental Electromagnetic Equations for Moving Bodies), 
co-authored by Albert Einstein and Jakob Laub, 
published in {\it Annalen der Physik}~\cite{EL1}. 
Courtesy of the {\it Leibniz-Institut f\"ur Astrophysik}, Potsdam, Germany.}
\label{EL}
\end{figure}

It is worth mentioning that Laub was a physicist who co-authored papers with Albert Einstein, 
about the Theory of Relativity, published in the journal {\it Annalen der Physik}, 
in 1908 and 1909 (cf. Fig.~\ref{EL}).~\cite{EL1, EL2, EL3, EL4}
~\footnote{Born as Jakub Laub, he emigrated to Argentina in 1911, and returned to Europe many years later.
I have found no evidence that Laub was aware of Einstein's effect to 
be measured during the total solar eclipse of 1912.}

Another commission from Argentina, 
from the La Plata Observatory, 
led by 
William Joseph Hussey (director), 
assisted by 
Henry Julius Colliau~\cite{Curtiss:1926}
and 
Bernhard Hildebrandt Dawson, 
arrived in Rio on October 2nd, in the 
ship {\it Arlanza}, coming from Buenos Aires. 
They visited the Brazilian National Observatory in the 
same day of their arrival. 
The La Plata Observatory team went to the city of Alfenas, 
in the state of Minas Gerais, together with  
the engineer Carlos Vieira Souto, 
from the Brazilian Railway Company, 
who acted as an interpreter of the foreign scientists. 
They decided to install their equipment in the courtyard 
of the city school.~\cite{OPaiz:12out1912}

\begin{figure}
\centering
\includegraphics[width=5.0in]{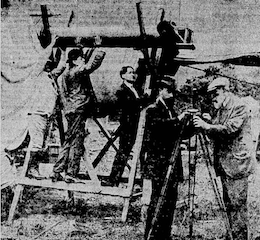}
\caption{Marc Ferrez (in the right) helping Milan Stefanik (in the left of Ferrez) 
during a rehearsal with a photometer~\cite{Epoca:12out1912}. 
In the back, the mounting of the Mailhat telescope of the French commission, 
with 10~m of focal distance, can also be seen. 
Courtesy of the {\it Funda\c{c}\~ao Biblioteca Nacional}, 
Rio de Janeiro, Brazil.}
\label{Ferrez}
\end{figure}

The Brazilian National Observatory organized two expeditions to 
observe the 1912 eclipse in different sites.
The main Brazilian team of the Brazilian Observatory, 
that went to Passa Quatro, was composed by
Henrique Morize (director), 
Domingos Fernandes da Costa (assistant astronomer), 
Mario Rodrigues de Souza (assistant astronomer), 
Alix Corr\^ea de Lemos (chief of the Meteorology section),
Alfredo de Castro e Almeida (mechanical), 
Augusto Soucasaux (photographer),
Gualter de Macedo Soares (auxiliar)
and, voluntarily,  
Ant\^onio Alves Ferreira da Silva 
(cf. Fig.~\ref{teams})~\cite{ANoite:01out1912p1}.~\footnote{
As a volunteer, went to Passa Quatro captain 
Ant\^onio Alves Ferreira da Silva
(accompanied by his wife), who had been 
previously in charge of the Time Service provided by 
the Brazilian National Observatory. In 1912, 
he was working for the Ministry of Foreign Affairs, 
as the sub-chief of the Brazilian government commission 
to determine the limits between Brazil and Bolivia,~\cite{JdC:10out1912} 
and obtained the necessary license to assist Morize 
during the eclpse observations.}
They were joined by the well-known photographer 
Marc Ferrez (cf. Fig.~\ref{teams}), who also helped with the preparations for the 
eclipse observation (cf. Fig.~\ref{Ferrez}).

\begin{figure}
\centering
\includegraphics[width=4.5in]{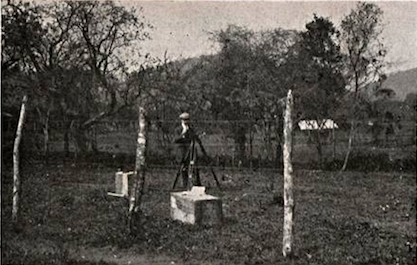}
\includegraphics[width=4.5in]{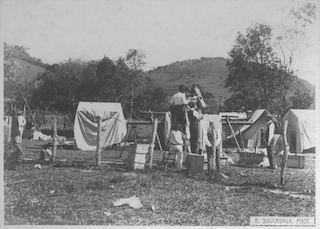}
\caption{
Top: Place chosen for the Brazilian settlement in  Passa Quatro, 
before the installation of the telescopes.~\cite{FonFon:12out1912p36}. 
Courtesy of the {\it Funda\c{c}\~ao Biblioteca Nacional}, 
Rio de Janeiro, Brazil.
Bottom: Brazilian settlement in  Passa Quatro. 
The team is adjusting the Steinhel equatorial telescope, under the 
supervision of Morize, standing at a distance, to the right.
The 8~m Mailhat telescope, 
brought from France by Stefanik, 
is seen in the background.
The Mailhat telescope is fed by a coelostat, 
to compensate Earth's rotation, 
which can be seen behind Morize. 
This Steinhel equatorial has a clock drive mechanism, 
and is coupled to a camera for 
9cmX12cm glass plates.
Courtesy of the {\it Biblioteca do Observat\'orio Nacional}, 
Rio de Janeiro, Brazil.}
\label{brazilian1}
\end{figure}

\begin{figure}
\centering
\includegraphics[width=4.9in]{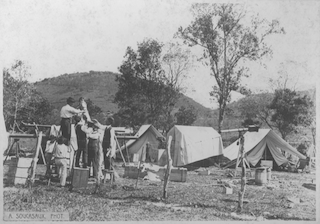}
\caption{Another view of the Brazilian settlement in Passa Quatro. 
Morize 
is standing close to the Steinheil equatorial telescope, 
which is being adjusted. 
In this image, apart from the Mailhat and Steinheil telescopes 
visible in Fig.~\ref{brazilian1}, a second 
Steinheil telescope
(positioned horizontally, without clock drive mechanism), 
coupled to a camera for 
18cmX24cm glass plates, can be seen, in the right. 
Courtesy of the {\it Biblioteca do Observat\'orio Nacional}, 
Rio de Janeiro, Brazil.}
\label{brazilian2}
\end{figure}

\begin{figure}
\centering
\includegraphics[width=5.0in]{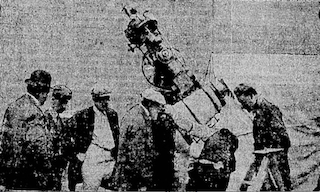}
\caption{Morize (in the right) aside of the Heyde equatorial telescope, 
in the Brazilian observation settlement, 
in Passa Quatro~\cite{Epoca:12out1912}.
M\'ario R. de Souza, Ant\^onio F. da Silva 
and Marc Ferrez 
can also be seen in this photo.
Courtesy of the {\it Funda\c{c}\~ao Biblioteca Nacional}, 
Rio de Janeiro, Brazil.}
\label{Heyde}
\end{figure}

\begin{figure}
\centering
\includegraphics[width=5.0in]{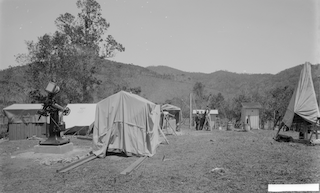}
\caption{One more view of the Brazilian settlement in Passa Quatro. 
In the left, the Heyde equatorial telescope appears uncovered. 
Covered with canvas, in the (foreground) center-left is the Mailhat telescope. 
In the (background) center-right, one of the Steinheil 
telescopes (the one without clock drive mechanism) can be seen, 
being adjusted by the Brazilian team. 
In the right, covered with canvas, is the main Steinheil telescope
(coupled to a clock drive mechanism). 
Courtesy of the {\it Biblioteca do Observat\'orio Nacional}, 
Rio de Janeiro, Brazil.}
\label{brazilian3}
\end{figure}

\begin{figure}
\centering
\includegraphics[width=4.9in]{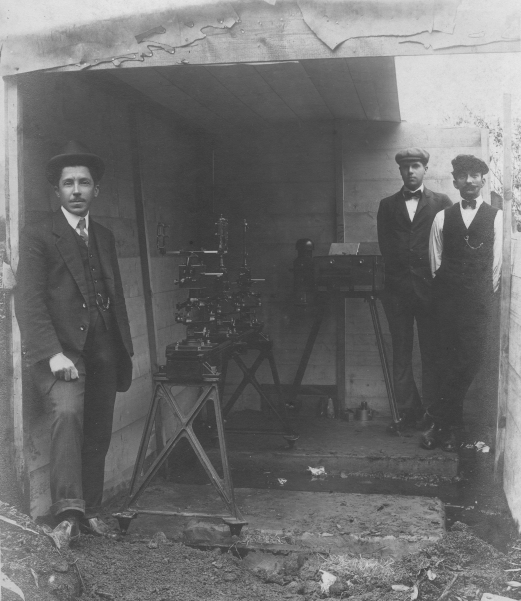}
\caption{Instruments used to measure Earth's magnetism, 
in Passa Quatro. 
Alix C. de Lemos is the first and Gualter de M. Soares
is the second, from left to right.
Courtesy of the {\it Biblioteca do Observat\'orio Nacional}, 
Rio de Janeiro, Brazil.}
\label{magnetism}
\end{figure}

Morize and some of his assistants made a prior trip to Passa Quatro, 
in the end of September for the installation of the instruments 
(cf. Figs.~\ref{brazilian1} and ~\ref{brazilian2})~\cite{ANoite:30set1912}.
\footnote{
Morize returned to Rio de Janeiro from this visit to Passa Quatro 
on October 1st.~\cite{JdB:01out1912}
}
The main Brazilian equipment were three telescopes, namely 
the Heyde equatorial (cf. Figs.~\ref{Heyde} and~\ref{brazilian3}), 
the 8~m Mailhat coupled to a coelostat (brought from France by Stefanik), 
and the Steinheil equatorial photoheliograph, with a double objective of 10~cm, 
and 1,50~m of focal distance 
(cf. Figs.~\ref{brazilian1} and~\ref{brazilian2}).~\cite{Inventory}~\footnote{
These same Mailhat and the Steinheil telescopes, 
which could not be properly tested during the 1912 eclipse, 
were brought by the Brazilian team to the city of Sobral 
(located in the Brazilian state of Cear\'a), years later, 
for the observation of the total solar eclipse of May 29th, 1919.
}
There was also a second Steinhel photoheliograph (cf. Fig.~\ref{brazilian2} and~\ref{brazilian3}) 
and a Bardou telescope. 
Other instruments were a Bendorf's electrometer
(Lord Kelvin type), 
to measure the air electric potential, 
variometers, to measure the magnetic declination (cf. Fig.~\ref{magnetism}), 
and an Adam Hilger spectrograph, to measure 
properties of the solar corona. 
Photometers, barographs, thermographs, hydrographs, 
anemometers, chronometers, etc., 
were also brought to Passa Quatro by the 
Brazilian team.~\cite{CdM:10out1912, Epoca:10out1912}

The operation of the telescopes was programmed as follows. 
Morize would take care mainly of the Heyde equatorial, 
assisted by 
Ant\^onio Ferreira da Silva.
The operation of the Mailhat telescope and coelostat 
was assigned to Mario de Souza.~\footnote{ 
During the tests, the Brazilian team was experiencing 
difficulties for the adjustment of the Mailhat telescope 
and there was the suspicion that it was 
defective.~\cite{CdM:10out1912}}
Domingos Costa would operate the main 
Steinheil equatorial photoheliograph.~\cite{CdM:10out1912} 
Alix C. de Lemos and Gualter de M. Soares were in charge of the 
measurements concerning Earth's magnetism (cf. Fig.~\ref{magnetism}).

\begin{figure}
\centering
\includegraphics[width=4.9in]{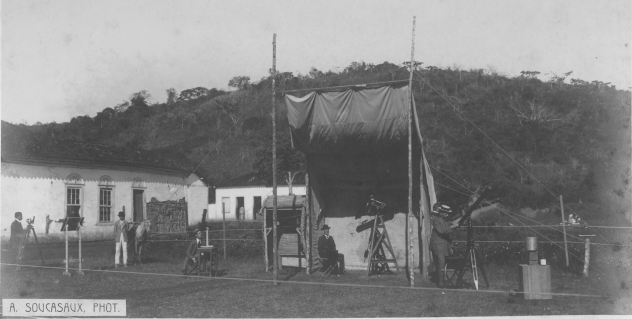}
\caption{Settlement of the Brazilian National 
Observatory at Silveiras, in the state of S\~ao Paulo.
Courtesy of the {\it Biblioteca do Observat\'orio Nacional}, 
Rio de Janeiro, Brazil.}
\label{Silveiras}
\end{figure}

The other commission of the Brazilian Observatory went to 
Silveiras, in the state of S\~ao Paulo (cf. Fig.~\ref{Silveiras}), 
led by Juli\~ao de Oliveira Lacaille (chief of the Astronomy section), 
together with 
Herminio F. Silva (assistant astronomer), 
Arthur de Castro Almeida (mechanical),
Valentim de Magalh\~aes (auxiliar) 
and 
Primo Flores (carpenter)~\cite{ANoite:01out1912p1, ANoite:30set1912}.

The Observatory of S\~ao Paulo, in Brazil, 
also sent an expedition, 
which was installed in a camp located 
in the west side of the city of Cruzeiro. 
This commission was led by 
Jos\'e Nunes Belfort de Mattos (director), 
with the assistance of 
Roberto Simon, 
Jos\'e Rangel Belfort de Mattos, 
Armando Fairbanks,
together with the mechanical 
Jacynto Schneck, among other people.
The instruments included 
a magnetometer and an actinometer, 
as well as a heliograph, one equatorial and two 
Bardou telescopes, anemometer, 
barometer, etc. 
In Cruzeiro, there was also 
a party from the 
Cinematography Brazilian 
International Company, 
which went there to register 
the eclipse.~\cite{CP:11out1912}
This S\~ao Paulo commission was joined by 
Rogerio Fajardo, teacher at the 
S\~ao Paulo Polytechnic School, together 
with his students, 
who intended to collaborate with the eclipse 
measurements.~\cite{JdC:10out1912, CP:11out1912}
\footnote{
There was also a team of priests and students, led by 
the jesuit Justino Maria Lombardi, from {\it Companhia de Jesus}; 
Jean Baptiste du Dr\'eneuf, rector of the School S\~ao Luiz, 
from the city of Itu; and
Vicente Prosperi, who thought Astronomy in the School Anchieta, 
from the city of Nova Friburgo.~\cite{CP:07out1912} 
} 

An investigation about the influence of the eclipse 
in radio-telegraphic
transmissions had also been organized 
by the Brazilian Government. 
For this investigation the Ministries of 
Transport 
(through the General Directorship 
of Telegraph)
and of the Navy were involved.
The engineer Leopoldo Ignacio Weiss and 
the telegraphist Manuel Soares Pinto Junior 
were in charge of this investigation.~\cite{JdB:09out1912}

\begin{figure}
\centering
\includegraphics[width=5.0in]{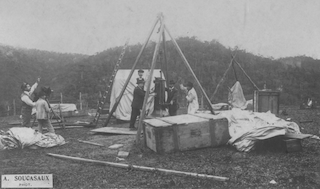}
\caption{Moving the base of the Heyde equatorial telescope, 
at the Brazilian settlement, in Passa Quatro. 
The photographer Augusto Soucasaux appears n the left, 
near his camera.
Courtesy of the {\it Biblioteca do Observat\'orio Nacional}, 
Rio de Janeiro, Brazil.}
\label{Heyde_base}
\end{figure}

Dozens of tons of equipment were transported by the Brazilian 
railway companies to the observation sites. 
Contributing to this heavy load, the base of the Heyde 
equatorial telescope of the Brazilian commission
(cf. Fig.~\ref{Heyde_base})
weighted around eight hundred kilograms.~\cite{Epoca:12out1912} 

Morize arrived back in Passa Quatro on October 5th, 
together with Ant\^onio F. da Silva 
and with the assistant Gualter M. Soares~\cite{ANoite:05out1912}.
During Morize's absence, the Brazilian party's 
arrangements in Passa Quatro were in charge of 
M\'ario R. de Souza~\cite{Eddington:26set1912}.
Domingos Costa had determined the exact time and 
latitude of the observation site in Passa Quatro.~\cite{CdM:10out1912}

\begin{figure}
\centering
\includegraphics[width=5.0in]{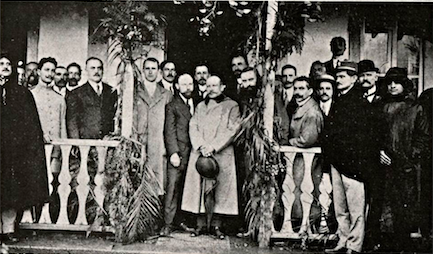}
\caption{Group of observers in Passa Quatro, 
gathered in the day of the 1912 eclipse.
The president of Brazil, Hermes da Fonseca, 
appears in the center of the photo, holding 
a hat and an umbrella, surrounded by the 
astronomers Stefanik, Eddington, and Morize,
among others~\cite{FonFon:19out1912p32}.
Courtesy of the {\it Funda\c{c}\~ao Biblioteca Nacional}, 
Rio de Janeiro, Brazil.}
\label{group_PQ}
\end{figure}

To observe the eclipse, even the president and 
the vice-president of Brazil, Hermes Rodrigues da Fonseca 
and Venceslau Br\'as Pereira Gomes, 
respectively, went to Passa Quatro, together with a 
group of politicians (some of them accompanied by their family),
including the Ministries of Foreign Affairs and 
of Finances, Lauro Muller and Francisco Salles, respectively, 
adding up to approximately 
eighty people~\cite{Epoca:11out1912}
 (cf. Fig.~\ref{group_PQ}). 
 Along the duration of the eclipse, they stayed 
 in the main house of the {\it Bella Vista} farm, 
 owned by Rodolpho Hess, where a nice lunch 
 was served afterwards. 
The estimations were that nearly a thousand people went to Passa Quatro 
to observe the phenomenon, 
including a docent of Astronomy (Gast\~ao Gomes) of the 
{\it Escola de Minas}, an engineering school located 
in the city of Ouro Preto (in the Sate of Minas Gerais), 
who brought some of his students to see the eclipse.~\cite{Epoca:11out1912}
The movie company {\it Brazil Films} even sent an operator 
to register the occasion in Passa Quatro. 

Stefanik and Kralicek, of the French expedition, 
were assisted in Passa Quatro by Gerald Waring and 
his wife, as well as 
by Barbosa Tigre. 
Eddington and Davidson, from the 
Greenwich expedition, were 
assisted by Atkinson, Aguirre and Andrews.
The amateur astronomer Worthington 
counted with the assistance of 
John Christopher Willis (director of the Rio de Janeiro 
Botanical Garden) and his wife, 
as well as Theophilus Lee.~\cite{CdM:10out1912} 

\begin{figure}
\centering
\includegraphics[width=5.0in]{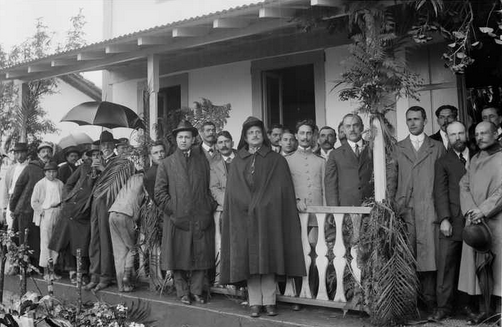}
\caption{The continuous rain did not allow the planed eclipse observations 
to take place in Passa Quatro on October 10, 1912.
Courtesy of the {\it Biblioteca do Observat\'orio Nacional}, 
Rio de Janeiro, Brazil.}
\label{rain}
\end{figure}

Unfortunately, heavy clouds covered the sky 
in Passa Quatro 
for a couple of days (cf. Fig.~\ref{rain}), 
including the whole eclipse duration.  
In the day of the eclipse the astronomers were ready, despite the rain, 
since there should be a hope that, at least during the totality, the portion 
of the sky with the eclipsed Sun could be seen, even if through the clouds. 
Unfortunately, this has not been the case in Passa Quatro, 
nor in any other observing station in Brazil.  
In Passa Quatro, the optical instruments were kept covered with canvas 
during the eclipse and, except for the 
measurements of the terrestrial magnetism, 
with the variometer, 
and of the atmospheric electricity, 
with the Bendorf's electrometer,~\cite{CdM:11out1912} 
performed by Domingos Costa; 
and the measurements performed with the polarimeter 
by Stefanik; basically no other instrumental operations  
could be done.
The camps remained muddy and 
 it was difficult even to walk around.~\cite{ANoite:10out1912p3}

The following estimations were published by the press: 
(i) first contact at 
08 hours, 56 minutes and 40 seconds (local time), 
(ii) maximum phase at 
10 hours, 16 minutes and 51 seconds, 
(iii) last contact at 
11 hours, 42 minutes and 54 seconds.~\cite{ANoite:10out1912p3}
According to Eddington: ``The rapid increase of the 
darkness a few seconds before totality was very striking, 
as was also the almost instantaneous brightening when it was over. 
Probably the state of the atmosphere caused the eclipse 
to be an unusually dark one.''~\cite{Eddington:1913}

Bad weather was faced in all the observation stations 
organized in Brazil. 
At Alfenas the weather was basically the same as in Passa Quatro. 
William Hussey, assisted by the other components of the 
Argentinean commission of the La Plata Observatory, tried to 
perform some light polarization measurements
during the eclipse.~\cite{JdC:11out1912} 

In the city of Cristina, where the astronomer Charles D. Perrine 
was leading the only team aiming to measure the light-bending effect 
predicted by Albert Einstein, the sky became completely cloudy 
two days before the eclipse. 
In the day before the eclipse the rain started and it continued 
raining without interruption until the day after the eclipse. 
The only possible measurements were of the light 
intensity through the clouds surrounding the eclipsed Sun,  
as well as an estimation of the beginning and end of the 
eclipse totality.~\cite{Perrine_Report_1912}
In the nearby Chilean Observatory camp, 
in {\it Boa Vista} farm, the same bad weather was faced.~\cite{AN_Ristenpart}

\begin{figure}
\centering
\includegraphics[width=5.0in]{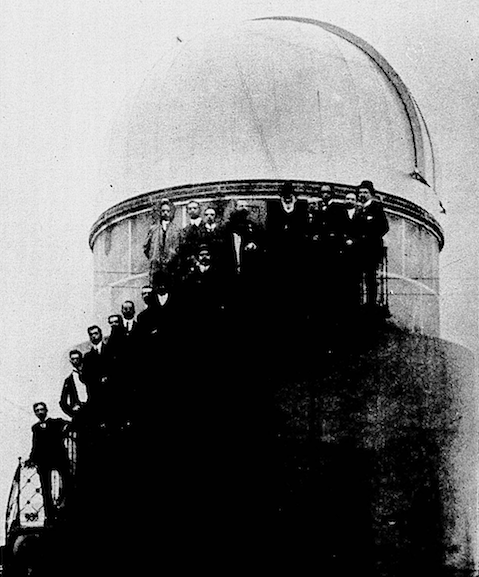}
\caption{People gathered 
around the dome of the equatorial telescope, 
in the headquarters of the 
Brazilian National 
Observatory, at {\it Morro do Castelo}, 
in the city of Rio de Janeiro, 
on October 10, 1912. 
~\cite{RdS:19out1912p15}
Courtesy of the {\it Funda\c{c}\~ao Biblioteca Nacional}, 
Rio de Janeiro, Brazil.}
\label{ON-RdS1}
\end{figure}

\begin{figure}
\centering
\includegraphics[width=5.0in]{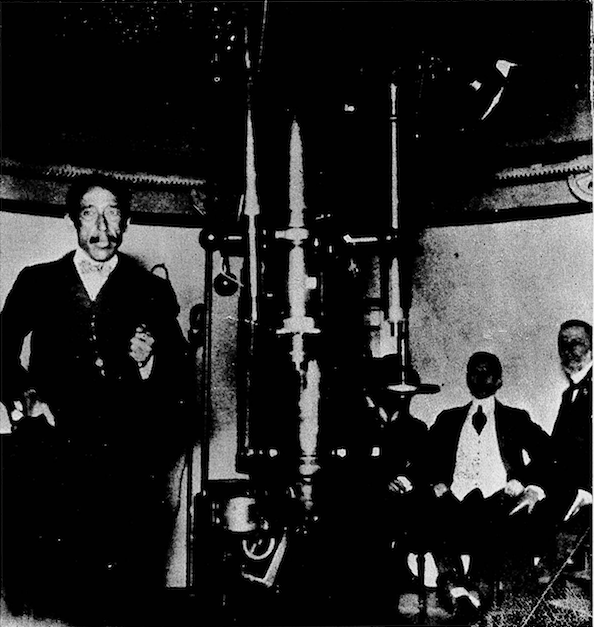}
\caption{Leopoldo Nery Voll\'u (left) aside of the 
equatorial telescope of the 
Brazilian National 
Observatory, at {\it Morro do Castelo}, 
in the day of the 1912 total solar eclipse. 
Brothero de Magalh\~aes and 
Mr. Ballardini 
appear standing in the right. 
~\cite{RdS:19out1912p15}
Courtesy of the {\it Funda\c{c}\~ao Biblioteca Nacional}, 
Rio de Janeiro, Brazil.}
\label{ON-RdS2}
\end{figure}

Although this 1912 eclipse of the Sun would only be visible as a partial one in 
the city of Rio de Janeiro, the Brazilian National 
Observatory planed observations at its headquarters, in {\it Morro do Castelo}, 
intending to use, for this purpose, 
the equatorial telescope available there (cf. Fig.~\ref{ON-RdS1}). 
The observations were in charge of 
Leopoldo Nery Voll\'u, 
assistant astronomer of the Observatory
 (cf. Fig.~\ref{ON-RdS2}). 
Other people present in the occasion were 
Macedo Soares, 
Sampaio Ferraz Filho, 
Luiz Rodrigues, 
Athanagildo Vilhena, 
Francisco Rodrigues de Souza, 
Eduardo Avilla, 
Francisco Goulart, 
Djalma Figueiredo, 
Galv\~ao Bueno, 
Raul Taunay, 
Euclydes Bomtempo, 
Alarico Milit\~ao, 
Emilio Loureiro e 
Alcides Carneiro~\cite{Epoca:11out1912}.
Due to the presence of heavy clouds 
during the whole duration the phenomenon, 
no observation of the (partially) eclipsed Sun could 
be performed in the city of Rio de Janeiro, 
although the time of the eclipse 
could be noticed by the people due to the 
corresponding darkness. 

\begin{figure}
\centering
\includegraphics[width=4.8in]{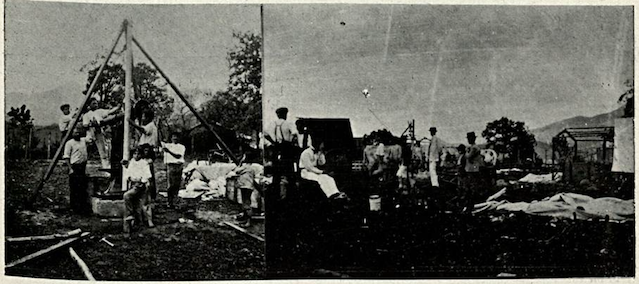}
\caption{Left: Dismounting the Mailhat coelostat 
of the Brazilian commission. 
Right: General view of the observation 
camp in Passa Quatro, during the 
dismounting.~\cite{FonFon:02nov1912p37}
Courtesy of the {\it Funda\c{c}\~ao Biblioteca Nacional}, 
Rio de Janeiro, Brazil.}
\label{desmontagem}
\end{figure}

\begin{figure}
\centering
\includegraphics[width=4.0in]{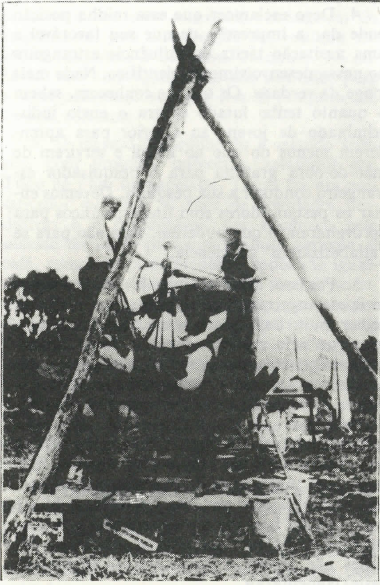}
\caption{Dismounting of the coelostat of the French 
commission.~\cite{Caffarelli:1980}.
This coelostat appears 
covered in the right of Fig.~\ref{french}.
Courtesy of {\it Ci\^encia e Cultura} magazine, 
Brazil.}
\label{dismount}
\end{figure}

Soon after the eclipse, 
the teams with settlements in the zone of the totality
started to pack their instruments to go back to 
their home institutions (cf. Figs.~\ref{desmontagem} and~\ref{dismount}). 
In the city of Passa Quatro, however, 
the Brazilian commission decided to 
keep for some days the equipment 
for air-electrical and magnetic measurements, 
in charge of 
Domingos Costa. 
Stefanik and Worthington 
also decided to stay for some 
more days in the camp. 
The motivation of Stefanik 
was to test the 
polarimeter of his invention.~\cite{ANoite:11out1912}
The French commission from the {\it Bureau des Longitudes} 
left Rio de Janeiro to Europe on October 23rd, 
on board of the ship {\it Atlantique}.~\cite{OPaiz:24out1912} 
\footnote{
In the minutes of the October 14th meeting of the 
 {\it Bureau des Longitudes} it is registered that 
Stefanik communicated that 
his observation of the solar eclipse in Brazil 
was hampered by the clouds, 
but nevertheless he could obtain some results.~\cite{BdL:14out1912}
}

Perrine, together with the C\'ordoba party, 
did a fast packing, leaving to Rio de Janeiro two days 
after the eclipse. 
They left Rio de Janeiro on October 15th in the ship {\it Asturias} to Buenos Aires, 
arriving on the 19th. 
Three days after that, they were back in C\'ordoba.~\cite{Perrine_Report_1912}

The Greenwich commission did not finish packing until October 16th, 
due to the wet weather, leaving Rio to Europe on October 23rd. 
Their packing-cases with the instruments did not reach Rio on time 
to go in the same boat as the commission, and were forwarded 
later, in another ship.~\cite{mnras73-0386}

\begin{figure}
\centering
\includegraphics[width=4.8in]{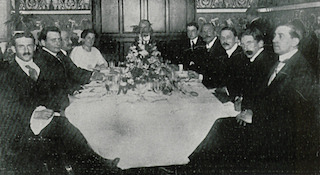}
\caption{Banquet in acknowledgement to 
Morize, offered by the Chilean representatives 
on October 15th, at the {\it Hotel dos Estrangeiros}, 
in Rio de Janeiro.~\cite{RdS:26out1912p14}
Courtesy of the {\it Funda\c{c}\~ao Biblioteca Nacional}, 
Rio de Janeiro, Brazil.}
\label{banquet}
\end{figure}

Back in Rio, Ristenpart gave a talk in the 
Commercial Museum ({\it Museu Comercial}), 
about the results obtained by 
the Chilean commission during 
the solar eclipse, to which even 
the Ministry of Foreign Affairs, 
Lauro Muller, has been invited.~\cite{JdC:17out1912}
~\footnote{
Jos\'e Belfort de Mattos, 
from the Observatory of S\~ao Paulo,
also gave a talk about the results 
obtained by his team during the eclipse 
observations, in a session of 
the S\~ao Paulo Scientific Society, 
on November 14th. 
Despite the cloudy weather 
at their observation site in Cruzeiro, 
he declared to have successfully 
investigated terrestrial magnetism, 
as well as 
actinometry.~\cite{CP:15nov1912} 
}
Some days before that, on October 15th, 
the Chilean authorities, in Rio, 
offered a banquet in homage to the director 
of the Brazilian National Observatory. 
Apart from Morize, the scientists Ristenpart, Knoche, 
Laub, accompanied by his wife, 
 were present in the occasion the Chilean 
 representatives in Brazil, 
 Samuel de Sousa Le\~ao Gracie (consul), 
 Alfredo Goycoolea Walton,
 and 
 Raul Cousi\~no Talavera, 
 as well as
 Frederico Sh\"afer
 and 
 Wilhelm Hipp, 
 the latter being the head of 
 Siemens Company in Rio
 (cf. Fig.~\ref{banquet})~\cite{JdC:17out1912, JdB:17out1912}.
On the 22nd of October, Ristenpart (accompanied by Grand\'on and W\"ust) 
left Rio de Janeiro to Buenos Aires, 
where he also gave a talk about the eclipse in Brazil.
From Argentina, the Santiago Observatory team went back 
to Chile.~\footnote{
Unfortunately, the ship {\it Oravia}, that was carrying the major part of 
the instruments of the Chilean commission in the return trip, 
sank near the Falkland Islands.~\cite{Grandon}
The Chilean team was not on board of the {\it Oravia}.
Friedrich Wilhelm Ristenpart killed himself on April 9th 1913, 
after a press campaign against him.
~\cite{Ristenpart, Duerbeck}
}
In December 1912, Ristenpart sent to publication from 
Santiago a report about the activities of the Chilean Observatory
team during the eclipse -- with special emphasis to the use 
of one of the selenium cells -- which was published in the following year in the journal 
{\it Astronomische Nachrichten}.~\cite{AN_Ristenpart}

Before leaving Brazil, 
Laub (with his wife), Knoche and Trollund 
went to the state of 
Esp\'irito Santo to visit the Native Americans, 
in the margins of the river {\it Doce}. 
They visited Colatina, where they met a 
group of {\it Krenak}.
They returned to Rio on October 28th, and in the same day they left, 
on board of the ship {\it Avon}, to 
Buenos Aires.~\cite{OPaiz:28out1912, AEpoca:28out1912}
In the following years, 
Laub and Knoche co-authored three 
publications about their results related to the  
meteorological and 
atmospheric electricity 
measurements performed at 
{\it Boa Vista} farm, in Cristina, 
during the 1912 total solar eclipse. 
~\cite{KL1, KL2, KL3}

\begin{figure}
\centering
\includegraphics[width=1.13in]{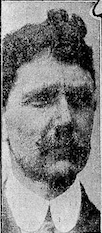}
\includegraphics[width=2.65in]{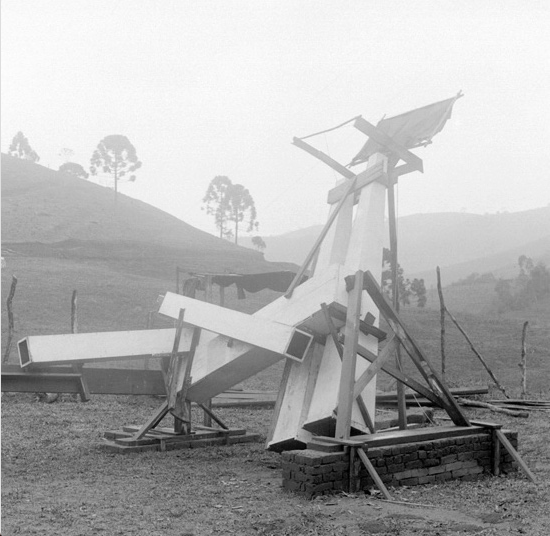}
\includegraphics[width=0.992in]{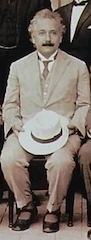}
\caption{Left: Charles Dillon Perrine,~\cite{ANoite:19set1912}  
the leader of the only team 
aiming to measure the light-deflection effect during the 1912 eclipse.
Courtesy of the {\it Funda\c{c}\~ao Biblioteca Nacional}, 
Rio de Janeiro, Brazil.
Center: Intramercurial cameras loaned by the Lick 
Observatory, and 
used by Perrine in Passa Quatro, in the attempt to 
measure light deflection by the Sun, during the 
October 10, 1912 eclipse.
Courtesy of the {\it Museo Astron\'omico del Observatorio Astron\'omico de C\'ordoba}, 
C\'ordoba, Argentina.
Right: Albert Einstein, who proposed in 1911 that light is deflected in a 
gravitational field, and that the gravitational light-bending 
by the Sun could be measured during a total solar eclipse.
Picture taken during Einstein's visit to the Brazilian National Observatory, in 1925.
Courtesy of {\it Museu de Astronomia e Ci\^encias Afins}, 
Rio de Janeiro, Brazil.}
\label{P_IC_E}
\end{figure}

We have reported about 
the expeditions organized to observe in Brazil 
the total solar eclipse of October 10, 1912. 
This has been one of the most expected eclipses visible in the Brazilian territory, 
but, unfortunately, due to the bad weather, it could not be properly appreciated by 
any of the observers.
Stands out the fact that it has been for this eclipse that the first expedition 
to verify 1911 Einstein's prediction of gravitational light bending has been 
organized and led by the American astronomer Charles D. Perrine, 
director of the Argentinian National Observatory, located in 
C\'ordoba (cf. Fig.~\ref{P_IC_E}).~\cite{Perrine1923, PM:2007, PPMCOG:2018, paolantonio2019, crisp_paola2020}
Curiously enough, Einstein's prediction made in 1911 was not correct, 
as Einstein himself realized and published in one of his 1915 works~\cite{einstein1915-03} 
that stablished the General Theory of Relativity.

The correct Einstein's prediction of light deflection by the Sun's 
gravitational field was finally verified in 1919, with the analysis 
of the total solar eclipse observations performed in Brazil, 
by Andrew Claude de la Cherois Crommelin and Davidson, 
and in Africa, by Eddington and Edwin Turner Cottingham. 
\footnote{
Curiously enough, together with the details published in the day of the 1912 eclipse, some of the 
newspapers in the city of Rio de Janeiro already announced the next total 
eclipse that could be visible in Brazil in 1919, mainly from the state of Cear\'a, 
whose totality would last for many more minutes 
than the 1912 one.~\cite{Epoca:10out1912, CdM:10out1912}
}
It is worth emphasizing the fact that Davidson and Eddington met 
in Rio de Janeiro with Perrine, in 1912. 
At that occasion, one of the main aims of Perrine expedition was to 
verify Einstein's effect of light bending, while the Greenwich expedition 
was essentially aiming to study properties of the solar corona, and, most 
probably, Eddington and Davidson were not aware of this 
Einstein's effect.

It is also worth noting that, to observe this 1912 eclipse, came to Brazil 
Jakob J. Laub, 
an earlier collaborator of Einstein in the Theory of Relativity. 
No evidence has been found that Laub new in 1912 about 
this Einstein's effect.

\section*{Acknowledgments}
I am grateful to 
(i) Santiago Paolantonio, from {\it Museo Astron\'omico del Observatorio Astron\'omico de C\'ordoba}, C\'ordoba, Argentina; 
(ii) Fred Espenak and Robert M. Candey, from NASA Goddard Space Flight Center, United States of America (USA); 
(iii) Luisa Haddad, Joop Rubens and Teresa Mora, from Special Collections and Archives, McHenry Library, University of California, Santa Cruz, USA; 
(iv) Shaun J. Hardy, from Carnegie Institution, Department of Terrestrial Magnetism, Washington, USA;
(v) James Kirwan, from Trinity College Library, Cambridge, United Kingdom (UK); 
(vi) Adam Perkins and Emma Saunders, 
from Cambridge University Library, Cambridge, UK;
(vii) Melissa Thies, from {\it Leibniz-Institut f\"ur Astrophysik}, Potsdam, Germany;
(viii) Emilie Kaftan, from {\it La Biblioth\`eque de l'Observatoire de Paris}, Paris, France;
(ix) Pascale Carpentier, from {\it Bureau des Longitudes}, Paris, France; 
(x) Michael Zeiler, from the Eclipse-Maps Website; 
(xi) Jorge Antonio Zanelli Iglesias, from {\it Centro de Estudios Cient\'ificos}, Valdivia, Chile;
(xii) David Az\'ocar, Mar\'ia Teresa Ruiz and Mario Andres Hamuy Wackenhut, from {\it Universidade de Chile}, Chile;
(xiii) Everaldo Pereira Frade, Luci Meri Guimar\~aes, Maria Celina Soares de Mello e Silva, and 
Maria Lucia de Niemeyer Matheus Loureiro, 
from {\it Museu de Astronomia e Ci\^encias Afins} (MAST), Rio de Janeiro, Brazil;  
(xiv) Carlos Henrique Veiga, Jailson Souza de Alcaniz, Jo\~ao Carlos Costa dos Anjos, 
K\'atia Teixeira dos Santos de Oliveira,
and
Teresinha de Jesus Alvarenga Rodrigues, 
from {\it Observat\'orio Nacional} (ON), Rio de Janeiro, Brazil; 
(xv) M\^onica Velloso Azevedo, from {\it Funda\c{c}\~ao Biblioteca Nacional}, 
Rio de Janeiro, Brazil; 
(xvi) Paulo Marques dos Santos, Rosa Maria Silva Santos and Sandra Aparecida Marques dos Santos, 
from {\it Instituto de Astronomia, Geof\'isica e Ci\^encias Atmosf\'ericas} 
of the {\it Universidade de S\~ao Paulo}, S\~ao Paulo, Brazil;
(xvii) Germana Fernandes Barata and Marcelo Knobel, from {\it Universidade de Campinas}, Campinas, Brazil;
and 
(xviii) Diene de C\'assia Barros de Oliveira, Leandro Fier Ribeiro and Thiago Kleber Costa dos Santos,  
from {\it Universidade Federal do Par\'a}, Bel\'em, Brazil.
I also acknowledge 
the partial financial support from 
{\it Conselho Nacional de Desenvolvimento Cient\'ifico e Tecnol\'ogico} (CNPq) 
and 
{\it Coordena\c{c}\~ao de Aperfei\c{c}oamento de Pessoal de N\'ivel Superior} (CAPES)
-- Finance Code 001, 
from Brazil.
This research has also received funding from the European Union's Horizon 2020 
research and innovation programme under the H2020-MSCA-RISE-2017 
Grant No. FunFiCO-777740.



\end{document}